\documentclass[epj]{svjour}
\usepackage{graphics}
\usepackage{epsfig}
\usepackage{amsmath}
\usepackage{longtable}
\begin{document}
\title{Measurement of the beam asymmetry in $\mathbf{\eta}$--photoproduction
       off the proton}
\author{\mail{H. Schmieden, Nussallee 12, 53115 Bonn, Germany,
        \email{schmieden@physik.uni-bonn.de}}
        D. Elsner\inst{1}, 
        A.V. Anisovich\inst{2,3},
        G. Anton\inst{4}, 
        J.C.S. Bacelar\inst{5},
        B. Bantes\inst{1},
        O. Bartholomy\inst{2},
        D. Bayadilov\inst{2,3},
        R. Beck\inst{2},
        Y.A. Beloglazov\inst{3},
        R. Bogend\"orfer\inst{4},
        R. Castelijns\inst{5},
        V. Crede\inst{2,6},
        H. Dutz\inst{1},
        A. Ehmanns\inst{2},
        K. Essig\inst{2},
        R. Ewald\inst{1},
        I. Fabry\inst{2},
        H. Flemming\inst{7},
        K. Fornet-Ponse\inst{1},
        M. Fuchs\inst{2},
        C. Funke\inst{2},
        R. Gothe\inst{1}\fnmsep\thanks{present address: University of South Carolina, USA},
        R. Gregor\inst{9},
        A.B. Gridnev\inst{3},
        E. Gutz\inst{2},
        S. H\"offgen\inst{1},
        P. Hoffmeister\inst{2},
        I. Horn\inst{2},
        J. H\"ossl\inst{4},
        I. Jaegle\inst{8},
        J. Junkersfeld\inst{2},
        H. Kalinowsky\inst{2},
        S. Kammer\inst{1},
        V. Kleber\inst{1},
        Frank Klein\inst{1},
        Friedrich Klein\inst{1},
        E. Klempt\inst{2},
        H. Koch\inst{7},
        M. Konrad\inst{1},
        B. Kopf\inst{7},
        M. Kotulla\inst{8,9},
        B. Krusche\inst{8},
        M. Lang\inst{2},
        J. Langheinrich\inst{1}\fnmsep\thanks{present address: University of South
        Carolina, USA},
        H. L\"ohner\inst{5}, 
        I.V. Lopatin\inst{3},
        J. Lotz\inst{2},
        S. Lugert\inst{9},
        H. Matth\"ay\inst{7},
        D. Menze\inst{1},
        T. Mertens\inst{8},
        J.G. Messchendorp\inst{5},
        V. Metag\inst{9},
        C. Morales\inst{1},
        M. Nanova\inst{9},
        V.A. Nikonov\inst{2,3},
        D. Novinski\inst{2,3},
        R. Novotny\inst{9},
        M. Ostrick\inst{1}\fnmsep\thanks{present address: University of Mainz,
        Germany},
        L.M. Pant\inst{9}\fnmsep\thanks{on leave from Nucl. Phys.
        Division, BARC, Mumbai, India},
        H. van Pee\inst{2,9},
        M. Pfeiffer\inst{9},
        A. Radkov\inst{3},
        A.V. Sarantsev\inst{2,3},
        S. Schadmand\inst{9}\fnmsep\thanks{present address: FZ J\"ulich, Germany},
        C. Schmidt\inst{2},
        H. Schmieden\inst{1},
        B. Schoch\inst{1},
        S. Shende\inst{5},
        G. Suft\inst{4},
        A. S{\"u}le\inst{1},
        V.V. Sumachev\inst{3},
        T. Szczepanek\inst{2},
        U. Thoma\inst{2,9},
        D. Trnka\inst{9},
        D. Walther\inst{1},
        C. Weinheimer\inst{2}\fnmsep\thanks{present address: University of
        M\"unster, Germany},
        \and C. Wendel\inst{2}
\newline(The CBELSA/TAPS collaboration)
}
\institute{Physikalisches Institut der Universit\"at Bonn, Germany 
           \and Helmholtz-Institut f\"ur Strahlen- und Kernphysik 
                der Universit\"at Bonn, Germany
           \and Petersburg Nuclear Physics Institute, Gatchina, Russia
           \and Physikalisches Institut, Universit\"at Erlangen, Germany
           \and KVI, University of Groningen, The Netherlands
           \and Department of Physics, Florida State University, Tallahassee,
           USA
           \and Physikalisches Institut, Ruhr-Universit\"at Bochum, Germany
           \and Physikalisches Institut, Universit\"at Basel, Switzerland
           \and II. Physikalisches Institut, Universit\"at Giessen, Germany
          }
\abstract{
The beam asymmetry, $\Sigma$, was measured at ELSA in the reaction
$\vec \gamma\, p \rightarrow \eta\, p$
using linearly polarised tagged photon beams,
produced by coherent bremsstrahlung off a diamond. 
The crystal was oriented to provide polarised
photons in the energy range $E_\gamma = 800$ to 1400 MeV
with the maximum polarisation of $P_\gamma = 49$\,\% obtained at
1305 MeV.
Both dominant decay modes of the $\eta$ into
two photons and $3\pi^0$ were used to
extract the beam asymmetry from the azimuthal modulation 
of the cross section.
The measurements cover the angular range 
$\Theta_\text{cm}\simeq 50$ -- 150 degrees.
Large asymmetries up to 80\,\% are observed, in agreement with a
previous measurement.
The \texttt{eta-MAID} model and the Bonn--Gatchina partial wave analysis 
describe the measurements,
but the required partial waves differ significantly.
\PACS{
      {13.60.-r}{Photon and charged-lepton interactions with hadrons} \and
      {13.60.Le}{Meson production} \and
      {13.88.+e}{Polarization in interactions and scattering} \and
      {14.20.Gk}{Baryon resonances with S=0}
     } 
} 
\authorrunning{D. Elsner et al.}
\titlerunning{Measurement of the beam asymmetry in 
              $\mathbf{\eta}$--photoproduction}
\maketitle
\section{Introduction}
\label{sec:intro} 

The rich excitation spectrum of the nucleon mirrors 
its complicated multi--quark inner dynamics. 
Therefore baryon spectroscopy is expected to provide 
benchmark data for any model of the nucleon, 
e.g. quark models in their variety \cite{CR00,Loering01} or,
increasingly in the near future,
Lattice QCD as an approximation of full Quantum Chromodynamics
\cite{KLW05}.
However, in many cases widths and density of states prohibit a clean 
identification, i.e. an unambiguous assignment of quantum numbers 
within a partial wave analysis.

The analyses are mostly based on pion and kaon induced reactions.
Since some excited states are suspected
to have a strongly disfavoured $\pi N$ coupling \cite{CR94},
photoinduced reactions offer a complementary access to the nucleon
spectrum, in particular in non-pionic final states.
This provided the motivation to search for expected (within quark models)  
but yet unobserved ``missing'' resonances in $\eta$ photoproduction
off the proton \cite{Crede05,SL02,Chen03}.

The $\eta$ channel provides a great simplification to the complex 
spectrum. 
Due to its isospin $I=0$
it only connects $N^*$ states ($I=1/2$)
to the nucleon ground state, but no $\Delta$ states ($I=3/2$). 
Nevertheless, an unambiguous extraction of all contributing partial waves
still requires a complete experiment with respect to the
reaction amplitudes.
Pseudoscalar meson photoproduction is determined by 4 complex amplitudes.
However, due to the inherent nonlinearities it is not sufficient to
measure $8-1(\text{overall phase})=7$ independent quantities, 
as could be naively expected.
Instead, it can  be shown that a minimum of 8 
observables needs to be measured \cite{CT97}.
Besides the differential cross section 
those include 3 single-spin and 4 double-spin observables.
The combination of double-spin observables can be appropriately
chosen, but cross section,
target asymmetry, $T$, recoil polarisation, $P$, and
beam asymmetry, $\Sigma$, are required in any case
(for a definition of the observables see e.g. ref.\cite{KDT95}).

Once a linearly polarised photon beam is provided,
the photon-beam asymmetry is already accessible
without polarised target or recoil polarimetry. 
For this case the cross section of pseudoscalar meson photoproduction
off a nucleon can be cast into the form \cite{KDT95}
\begin{equation}
\frac{d\sigma}{d\Omega} = \frac{d\sigma_0}{d\Omega}\:
                          \left( 1 + P_\gamma\,\Sigma\,\cos 2\Phi \right),
\label{eq:xsec}
\end{equation}
where $\sigma_0$ denotes the polarisation independent cross section,
$P_\gamma$ the degree of linear polarisation of the incident photon beam,
and $\Phi$ the azimuthal orientation of the reaction plane 
with respect to the plane of linear polarisation.
While in principle it suffices to determine $d\sigma/d\Omega$ around
$\Phi=0$ and $\Phi=90$\,degrees, it is more favourable to
extract the beam asymmetry from the modulation 
of the cross section over the full azimuthal circle,
since systematic effects are better under control.
Thus, a cylindrically symmetric detector such as \texttt{Crystal Barrel}
\cite{CBarrel} is particularly suited to measure 
$\Sigma$ in $\eta$ photoproduction.

Most previous experiments investigated differential cross sections
\cite{Crede05,Krusche95,Renard02,Dugger02}. 
But there are also a few measurements of single polarisation observables.
Heusch et al. \cite{Heusch70} determined the recoil proton polarisation
in $\eta$ photoproduction between $0.8$ GeV and $1.1$ GeV in a spark chamber
experiment.
The target asymmetry was measured at the Bonn synchrotron \cite{Bock81}.
A first measurement of the photon beam asymmetry using 
linearly polarised photon beams was accomplished at the 
laser backscattering facility GRAAL at the ESRF Grenoble \cite{Ajaka98}. 
The GRAAL experiments were later on extended to higher energies and
preliminary results have been presented at conferences \cite{Kouznetsov02}.
Large $\Sigma \simeq 0.5$ were obtained in the near-threshold region.
Contrary, in $\eta$ electroproduction the $TT$ interference cross section, 
which is related to $\Sigma$, was
found consistent with zero over almost all the range in 
$Q^2=0.25$ --- $1.5$ GeV/c$^2$ in the threshold region \cite{Thompson01}.

In order to clarify this situation and to extend the energy range in
$\eta$ photoproduction we carried out experiments with linearly polarised
tagged photon beams at the electron accelerator ELSA \cite{Hillert06} 
of the University of Bonn.
The following section is first devoted to the experimental setup.
In addition to the basic analysis steps, section\,\ref{sec:analysis} 
then describes the method of extracting $\Sigma$.
The results are discussed in section \ref{sec:results} and,
after a brief summary, 
tabulated in the appendix.

\section{Experimental Setup}
\label{sec:1}

\begin{figure}
\resizebox{0.97\columnwidth}{!}{%
  \includegraphics{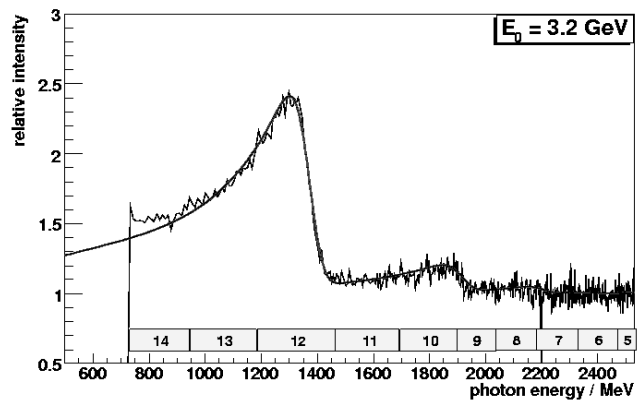} }
\caption{The measured coherent bremsstrahlung intensity 
         normalised to an incoherent spectrum (histogram, see text) 
         in comparison to an improved version \cite{Elsner06}
         of the ANB-calculation \cite{ANB} (full curve).
         The diamond
         radiator was set for an intensity maximum at $E_\gamma=1305$\,MeV.
         The numbered blocks indicate the ranges covered by the 14 timing 
         scintillators of the tagging detector.}
\label{fig:coh_spect}       
\end{figure}

Electron beams of $E_0 = 3.2$ GeV were used to produce
coherent bremsstrahlung from a $500\,\mu$m thick diamond crystal.
Electrons which radiated a photon are momentum analysed using
a magnetic dipole (tagging-) spectrometer.
Its detection system consists of 14 plastic scintillators providing fast
timing and additional hodoscopes to achieve the required energy resolution:
The range of low electron energies, 
corresponding to $E_\gamma = 0.8 ... 0.92\,E_0$,
is covered by a multi-wire proportional chamber,
a 480 channel double-layer scintillating fibre detector complements the range
$0.18 ... 0.8\,E_0$.
At the nominal setting of $E_0 = 3.2$\,GeV
the energy resolution varies between 2\,MeV for the high photon energies
and 25\,MeV for the low energies.
Since the photon beam remained virtually uncollimated, the measured
electron spectrum directly reflects the photon spectrum.
\begin{figure}
\resizebox{0.97\columnwidth}{!}{%
  \includegraphics{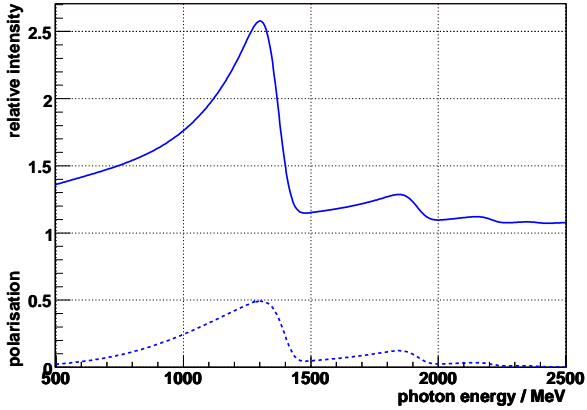} }
\caption{Calculation of relative bremsstrahlung intensity (top curve) and 
         corresponding degree of linear polarisation (bottom curve)
         using an improved version \cite{Elsner06} of
         the ANB bremsstrahlung code \cite{ANB}
         with scaled incoherent contribution 
         (see text).
         A scaling factor of $1.35$ is used to achieve 
         best agreement with the measured spectra 
         (cf. Fig.\,\protect{\ref{fig:coh_spect}}).}
\label{fig:photpol}       
\end{figure}

Fig.\ref{fig:coh_spect} shows the photon energy distribution obtained 
from the diamond radiator, measured through the detection of the 
corresponding electrons in the tagging system.
This spectrum is
normalised to the spectrum of an amorphous copper radiator.
Hence, a constant run of the curve 
corresponds to the ordinary $\sim 1/E_\gamma$  
dependence in the bremsstrahlung process.
This representation accentuates the coherence effect,
which manifests itself in clear peaks. 
Within the range of the coherent peaks the bremsstrahlung recoil is 
transferred to the whole crystal as opposed to individual nuclei in the
incoherent process,
thus fixing the plane of electron deflection very tightly relative to the 
orientation of the crystal lattice.
Consequently, the emitted photons are linearly polarised \cite{Timm69}.
The maximum achievable degree of polarisation decreases with increasing
photon energy, $P_\gamma \simeq 0.4$ is obtained at $E_\gamma=E_0/2$.
The orientation of the linear polarisation and the 
position of the maximum in the photon energy-spectrum 
can be deliberately chosen
through appropriate alignment of the crystal relative to the electron
beam direction.
We used a crystal setting to obtain the polarisation maximum at 1305
MeV.  
Vertical orientation of the polarisation vector was chosen, since
the vertical emittance of the electron beam is about an order of magnitude
better than in horizontal direction.  
A dedicated commercial 5-axis goniometer\footnote{{\it Newport} company}
enabled the accurate crystal alignment with typical angular uncertainties
of $\delta < 170\,\mu$rad.

The curve in Fig.\ref{fig:coh_spect} represents 
a calculation of the spectrum using an improved version \cite{Elsner06} of 
the original ANB 
(``analytic bremsstrahlung calculation'') software \cite{ANB} 
from T\"ubingen University.
It nicely describes the measured spectrum. 
This level of agreement can be only obtained, if the {\it incoherent}
part of the ANB calculation is scaled by a factor of $1.35$.
This was traced back to 
an inaccurate inclusion of multiple scattering
and an uncertainty in the atomic form factors \cite{Elsner06}.
Using the form factor parametrisation after Schiff \cite{Sch51} instead
that of Hubbell \cite{Hub59} improves the agreement significantly.

The relative strengths of coherent and incoherent contributions
determine the absolute value of linear polarisation.
It can be obtained from {\it any} fit of the spectrum as long 
as there is no overlap of different reciprocal lattice vectors 
--- which can correspond to different orientations of the resulting
    polarisation vector ---
within a given energy interval. 
This condition is surely fulfilled, if adjacent peak regions do not overlap. 
In this respect the mentioned re-scaling of the incoherent contributions
introduces no significant error. 
As can be seen from Fig.\ref{fig:coh_spect},
in our particular case there is only a tiny overlap between the adjacent
peaks.
Furthermore, both of them even result in the same orientation
of the polarisation vector.

Fig.\ref{fig:photpol} shows the ANB-calculated 
relative photon intensity spectrum in conjunction with 
the calculated photon polarisation.
The maximum polarisation of $P_\gamma = 0.49$ is obtained at 
$E_\gamma = 1305$ MeV, as expected.
An absolute error of $\delta P_\gamma < 0.02$ is estimated.
The total photon flux was up to $2 \times 10^7$ s$^{-1}$. 

\begin{figure}
\begin{center}
\resizebox{0.97\columnwidth}{!}{%
  \includegraphics{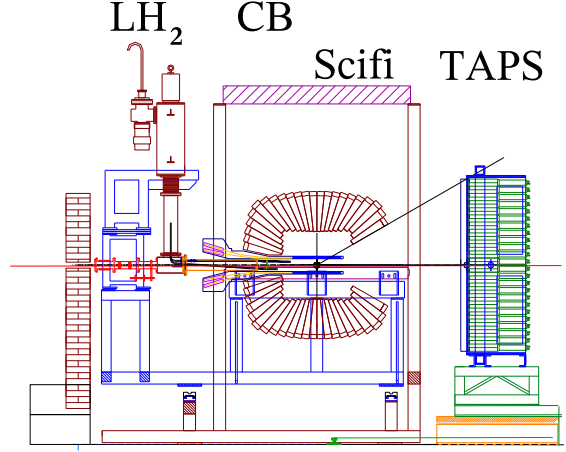} }
\end{center}
\caption{Setup of the detector system as described in the text.
         The photon beam enters from left.}
\label{fig:setup}       
\end{figure}

The detector setup of the experiment is depicted in Fig.\ref{fig:setup}.
The linearly polarised photon beam was incident on a $5.3$ cm long 
liquid hydrogen target with 80\,$\mu$m Kapton windows
\cite{Kop02}.
A three layer scintillating fibre detector \cite{Suft05} 
surrounded the target within the polar angular range from 
15 to 165 degrees. 
It determined a piercing point for charged particles.
Both, charged particles and photons were detected in the 
\texttt{Crystal Barrel} detector \cite{CBarrel}. 
It was cylindrically arranged around the target
with 1290 individual CsI(Tl) crystals in 23 rings, 
covering a polar angular range of 30 --- 168 degrees.
The crystals of 16 radiation lengths guaranteed nearly full longitudinal 
shower containment. 
In transverse direction electromagnetic showers extended over up to 
30 modules.
For photons an energy resolution of 
$\sigma_{E_\gamma}/E_\gamma 
= 2.5\,\%/^4\sqrt{E_\gamma/\text{GeV}}$
and an angular resolution of $\sigma_{\Theta,\Phi} \simeq 1.1$\,degree 
was obtained.

The $5.8$ --- 30 degree forward cone was covered by the 
\texttt{TAPS} detector \cite{TAPS},
set up in one hexagonally shaped wall of 528 BaF$_2$ modules.
For photons between 45 and 790 MeV the energy resolution is
$\sigma_{E_\gamma}/E_\gamma 
= \left(0.59/\sqrt{E_\gamma/\text{GeV}}+1.9\right)\%$
\cite{Gabler94}.
The position of photon incidence could be resolved within
20\,mm.
For charged particle recognition
each \texttt{TAPS} module has a 5\,mm plastic scintillator in front of it.
\begin{figure*}
\begin{center}
\resizebox{0.8\textwidth}{!}{%
  \includegraphics{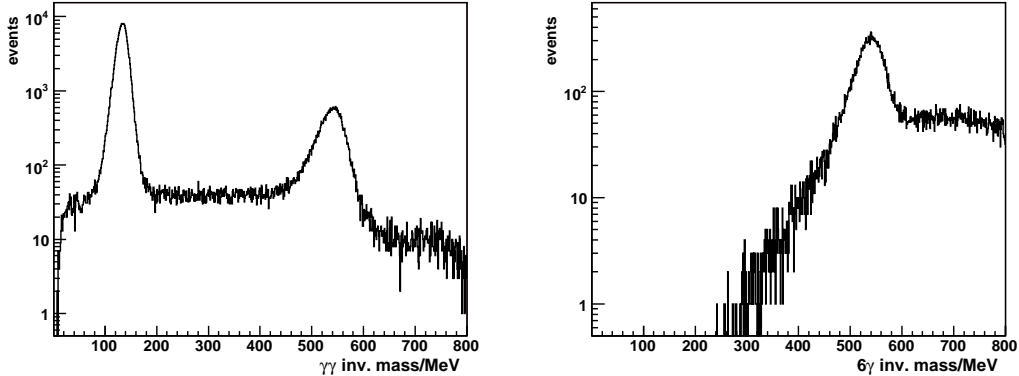} }
\end{center}
\caption{Invariant mass distribution 
         after standard kinematic analysis cuts.
         Left: Two photon invariant mass distribution 
         for the 3--cluster data set; 
         signal widths of $\sigma_{\pi^0} = 10$ MeV and 
         $\sigma_{\eta} = 22$ MeV are obtained.
         Right: 6 photon invariant mass distribution 
         for the 7--cluster data set with
         $\sigma_{\eta} = 25$ MeV.
         Note the logarithmic scale.}
\label{fig:inv_mass}       
\end{figure*}

In contrast to \texttt{Crystal Barrel}, the fast \texttt{TAPS} detectors are 
individually equipped with photomultiplier readout.
Thus, the first level trigger was derived from \texttt{TAPS}, requiring
either $\geq 2$ hits above a low threshold ($A$) or, alternatively,
$\geq 1$ hit above a high threshold ($B$). 
Using, within $\simeq 10\,\mu$s, 
a fast cluster recognition \cite{Flemming00} for the \texttt{Crystal Barrel} 
as second level trigger ($C$), 
the total trigger condition required 
$[A \lor (B \land C)]$, 
with 2 clusters 
identified at second level.

\section{Event reconstruction and data analysis}
\label{sec:analysis}

To enrich the $\eta\,p$ final state,
the occurence of, in total, either three or seven detector hits was required
during the offline analysis, 
corresponding to two or six photons and the proton.
In particular photon hits usually fire a cluster of adjacent 
crystals whose energy is summed over.
After the basic detector calibrations from the data itself,
the $\eta$ meson is identified in either of its major decay modes
into two photons or $3\pi^0$.
Fig.\,\ref{fig:inv_mass} shows the respective invariant mass
distributions, obtained after
only basic kinematic cuts have been applied in order to ensure
consistence of the azimuthal angles (i.e. coplanarity) 
and polar angles involved.
No cuts were applied on the energy of the respective hit of 
the proton candidate.
The signal widths in Fig. \ref{fig:inv_mass} are 
$\sigma_{\eta\rightarrow\gamma\gamma} = 22$\,MeV and
$\sigma_{\eta\rightarrow 3\pi^0} = 25$\,MeV, respectively.
To avoid any possible bias from detector inefficiencies 
on the azimuthal distributions,
the proton was {\it not} positively identified by using 
the signals of the inner scintillating fibre detector of 
the barrel or the veto detectors of \texttt{TAPS}.
Instead, all combinatorial possibilities were processed,
i.e. 3 for the 3--cluster events and 21 for the 7--cluster events.
A cut on the missing mass applied to the proton candidates subsequently 
yielded a clean separation.  
No kinematic fit was used to improve the separation,
nor to increase the resolution.

As can be seen from Fig.\,\ref{fig:inv_mass}, the background below
the $\eta$ peaks is very small (note the logarithmic scale).
It varies with photon energy and thus was determined in each
bin of $E_\gamma$.
Two different fits were used to interpolate the background between
the edges of the signal, linear and gaussian.
From the difference the possible systematic error was estimated
which may be due to the background subtraction scheme.

\subsection{Beam asymmetry}
\label{sec:asymmetry}

Cuts of $3 \sigma$ widhts around the $\eta$--mass 
in the invariant mass spectra 
(Fig.\,\ref{fig:inv_mass}) yielded a clean event sample.
To extract the photon beam asymmetry according to Eq.\,\ref{eq:xsec},
a fit of the azimuthal event distribution was performed:
\begin{equation}
f(\Phi) = A + B\,\cos (2\Phi).
\label{eq:fit}
\end{equation}
\begin{figure} 
 \resizebox{0.97\columnwidth}{!}{%
 \includegraphics{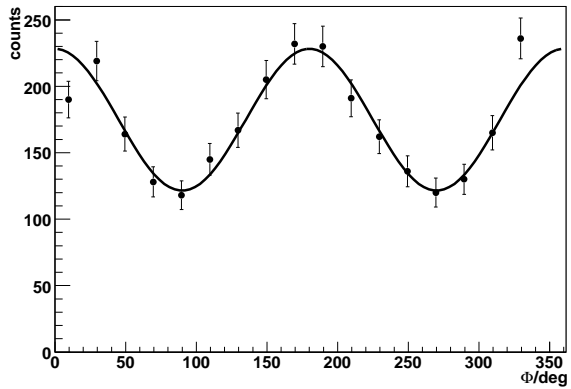} }
\caption{Example of a measured $\Phi$ distribution in the bin
         $E_\gamma = 1240$ --- $1350$ MeV and 
         $\Theta_\text{cm} = 66$ --- 92 degrees for the 
         $\eta \rightarrow 2\gamma$ decay channel.
         The event-weighted average polarisation was $P_\gamma = 47.3\,\%$.
         }
\label{fig:Phi_distr}       
\end{figure}
An example for one bin in $E_\gamma$ and $\Theta^{cm}$ is shown in 
Fig.\,\ref{fig:Phi_distr}.
The ratio $B/A$ of the fit determines the product of beam asymmetry and
photon polarisation, $P_\gamma \Sigma$, of Eq.\,\ref{eq:xsec}.
Since there is a strict relation between the photon energy 
and the photon polarisation (c.f. Fig.\,\ref{fig:photpol}), and
the appropriate photon energy can be assigned to each single event, 
it is possible to determine the event-weighted average polarisation
in each bin of photon energy.

\begin{figure*}
\begin{center}
\epsfig{file=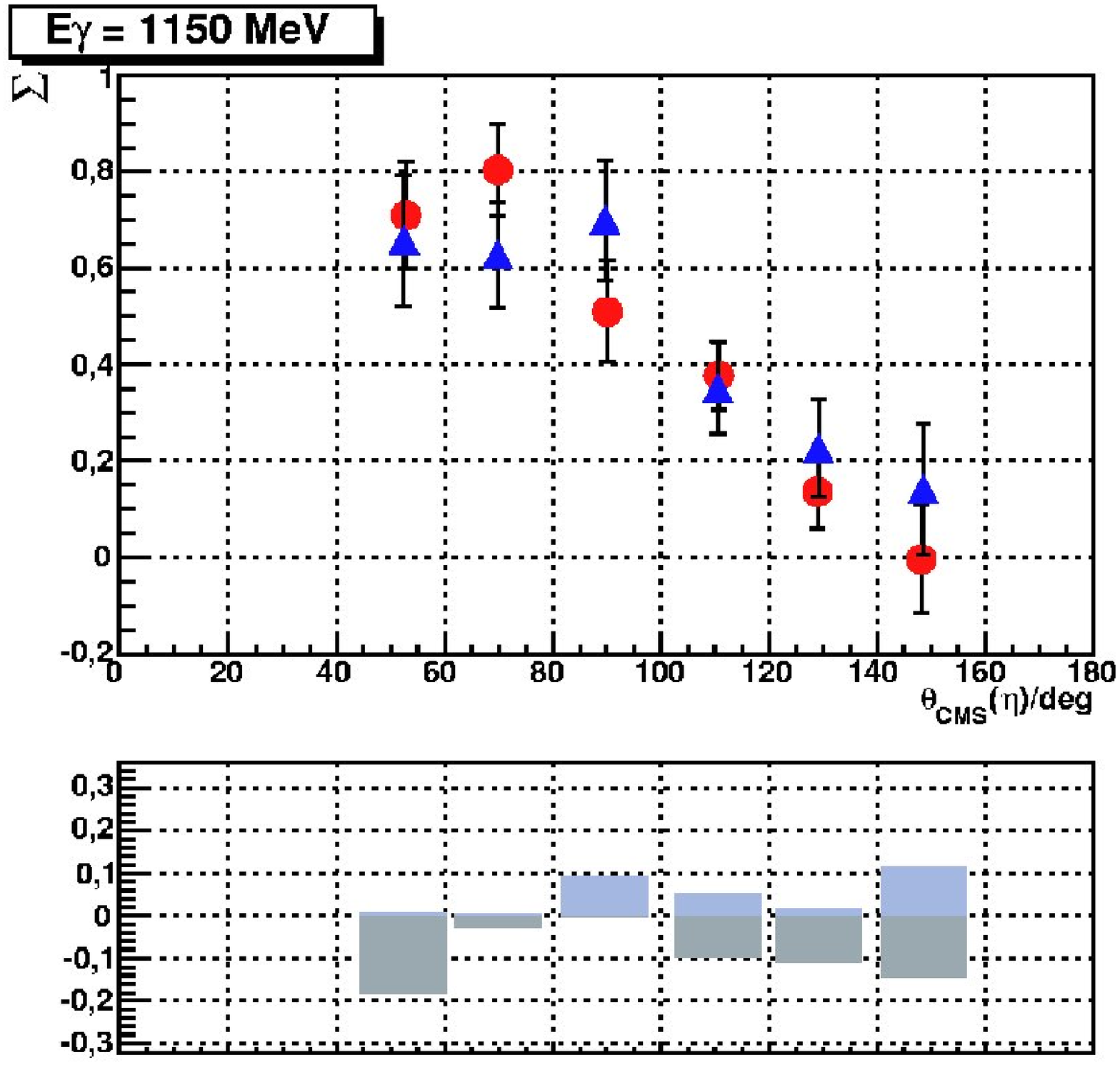,width=7.3cm,angle=-90}
\hspace{0.8cm}
\epsfig{file=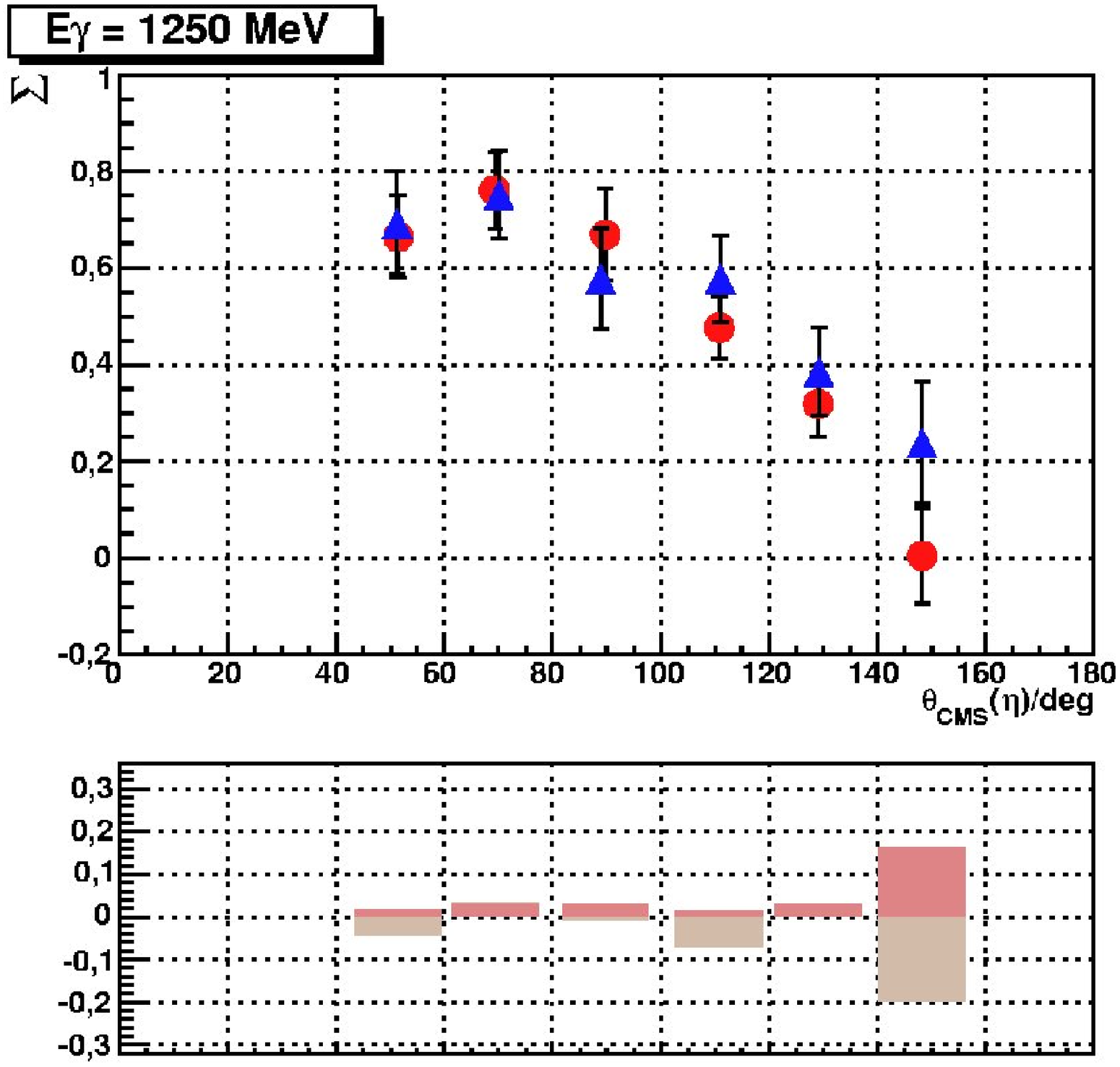,width=7.3cm,angle=-90}
\end{center}
\caption{Measured photon asymmetry, $\Sigma$, as extracted from the
         decay channels 
         $\eta \rightarrow 2\,\gamma$ (dots) and
         $\eta \rightarrow 3\,\pi^0$ (triangles) for the two 
         photon energy bins 1150 MeV (left) and 1250 MeV (right).
         The bar charts indicate the total fluctuation 
         ({\em no} 1--$\sigma$ errors) of $\Sigma$ if
         extracted from the $\Phi$ ranges 0 --- 180 degrees (light)
         and 180 --- 360 degrees (dark) seperately,
         instead of using the full range.
         Bottom left is for the 
         $\eta \rightarrow 3\pi^0$ channel in the 1150 MeV bin, 
         bottom right for the $\eta \rightarrow 2\gamma$ mode
         in the 1250 MeV bin.}
\label{fig:Sigma_2g-3pi}       
\end{figure*}

The photon asymmetries extracted from the $\eta \rightarrow 2\gamma$ and the 
$\eta \rightarrow 3\pi^0$ decay channels agree very well.
This is illustrated in Fig.\,\ref{fig:Sigma_2g-3pi} (top) where,
as examples, the two photon energy bins $1150 \pm 50$ MeV (left) 
and $1250 \pm 50$ MeV (right) are shown.

In order to detect possible false detector asymmetries, 
the uniformity of the event distribution of the laboratory angles  
$\Theta$ versus $\Phi$ was routinely inspected
\cite{Elsner06}.
Most detected problems could be removed in the offline
analysis.
Other sources of false asymmetries were identified
but could not be completely remedied, 
e.g. trigger inefficiencies within certain angular regions.
In such bins the corresponding $\Phi$--regions were excluded 
from the fit of Eq.\,\ref{eq:fit}.
The remaining systematic error is estimated through the difference
of separate fits of the 0 --- 180 and 180 --- 360 degree azimuthal 
regions to the full fit. 
The differences are shown as the bar graphs on the bottom of 
Fig.\,\ref{fig:Sigma_2g-3pi},
left in the $2\gamma$, right in the $3\pi^0$ 
decay of the $\eta$ meson.
Note that these estimates are correlated with the 
statistical errors. 
%

It turned out that the 
angle dependent inefficiencies
provide by far the major contribution to the systematic error of 
this experiment.
In contrast, the remaining uncertainty of the beam polarisation
affects the final result much less, and the effect of the background 
subtraction is almost negligible. 
The total error remains, however, still dominated by statistics 
as can also be seen from the table of results in the appendix.

\section{Results and discussion}
\label{sec:results}

\begin{figure*}
\begin{center}
\resizebox{0.8\textwidth}{!}{%
  \includegraphics{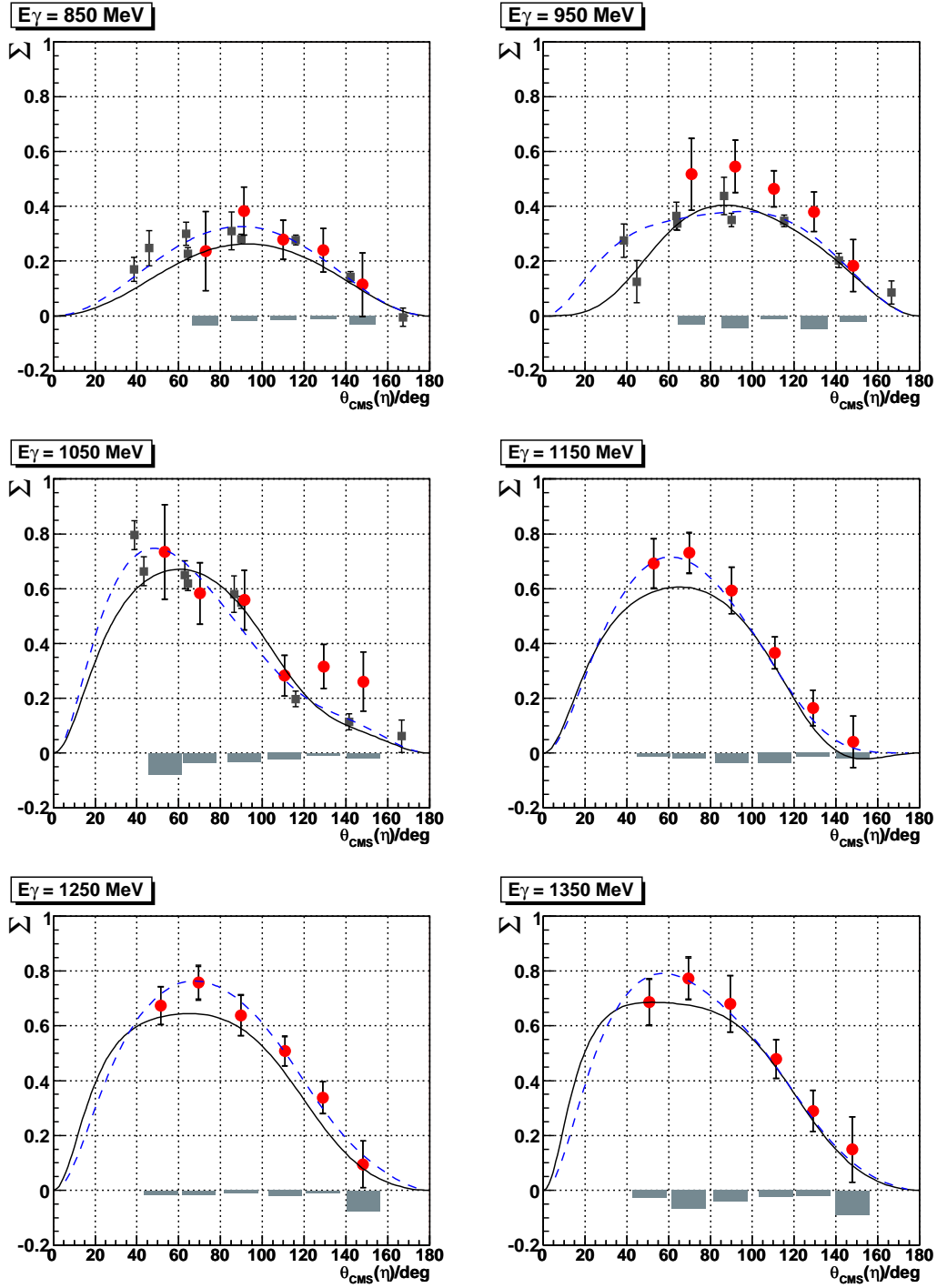} }
\end{center}
\caption{Photon asymmetry from the combined $\eta$ decay modes
         (filled circles)
         with statistical errors. The systematical error is indicated
         by the bar chart.
         Our results 
         are compared to the published data (boxes) of the 
         GRAAL collaboration \cite{Ajaka98} (see also text).
         The curves represent calculations of \texttt{eta-MAID}
         \cite{CYTD02} (full) and the Bonn--Gatchina partial wave analysis 
         \texttt{BnGa} \cite{Anisovich05} (dashed).}
\label{fig:Eta_all}       
\end{figure*}

The combined results of the $\eta \rightarrow 2\gamma$ and
$\eta \rightarrow 3\pi^0$ data sets are presented in 
Fig.\,\ref{fig:Eta_all}.
Statistical errors are directly attached to the data points.
Since determined from the $\chi^2$ of the fit of Eq. \ref{eq:fit},
these statistical errors may still carry some correlation to
systematics.
The estimated total systematic uncertainty is indicated by the bars.

Nice agreement is found with the published GRAAL data of Ajaka et al.
\cite{Ajaka98}.
This provides confidence that the analysis chain is well under control
on the level of the presented errors, 
in particular the determination of the degree of linear polarisation
and the extraction of the azimuthal asymmetries, 
the latter despite the fact that, due to the unfavourable horizontal
beam emittance, no data were taken with the polarisation plane rotated
by 90 degrees, as was done by the GRAAL collaboration.
More recent but yet preliminary 
(and hence here not shown)
data of the GRAAL collaboration,
extended in energy up to 1445 MeV \cite{Kouznetsov02},
do also nicely agree with our data.

In Fig.\ref{fig:Eta_all} our new data are 
compared to two standard calculations, 
the Mainz isobar model \texttt{eta-MAID} \cite{CYTD02}  
and the Bonn--Gatchina partial wave analysis
\texttt{BnGa} \cite{Anisovich05}.
\begin{figure*}
\epsfig{file=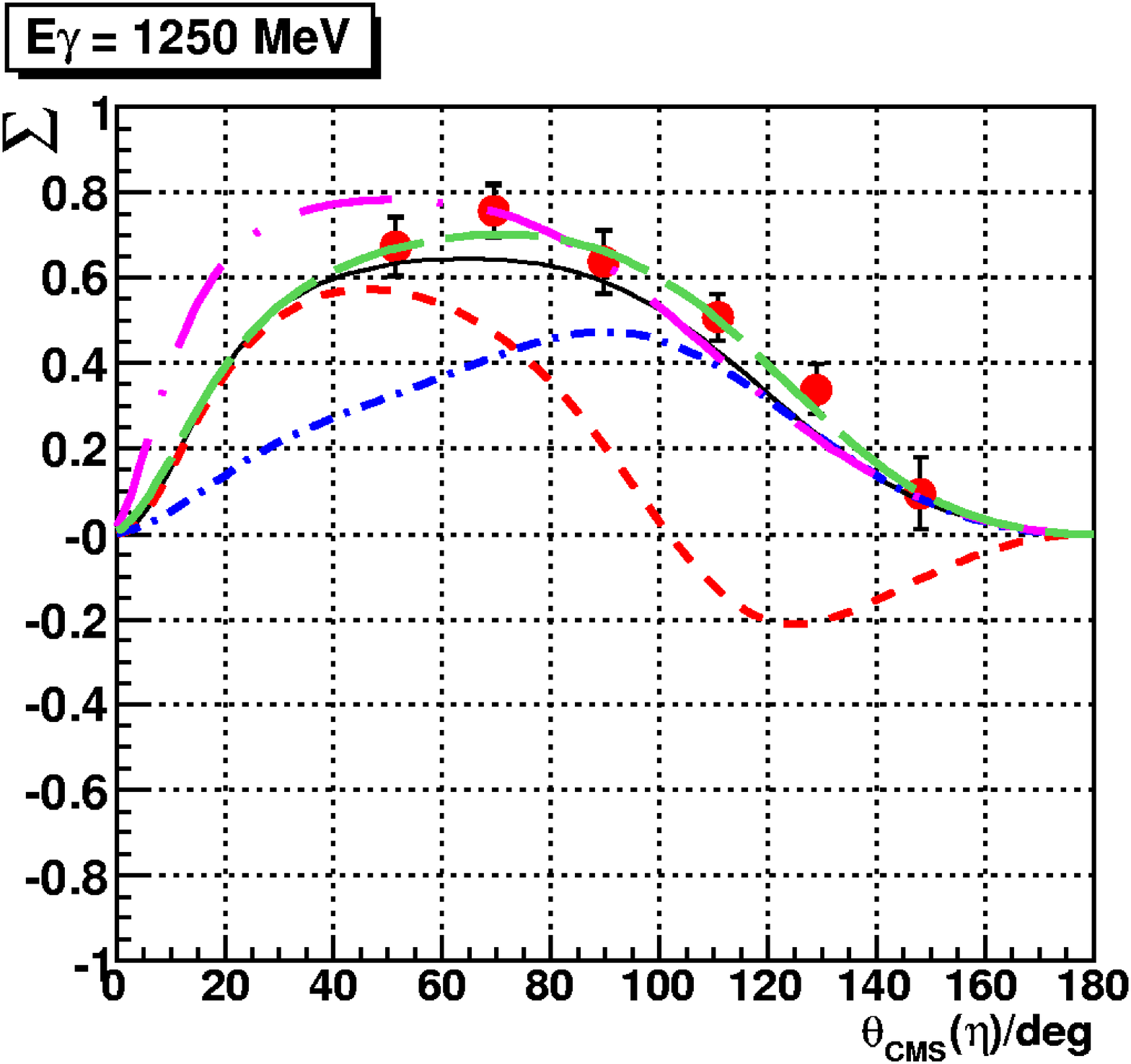,width=7.5cm,angle=0}
\hspace{1.5cm}
\epsfig{file=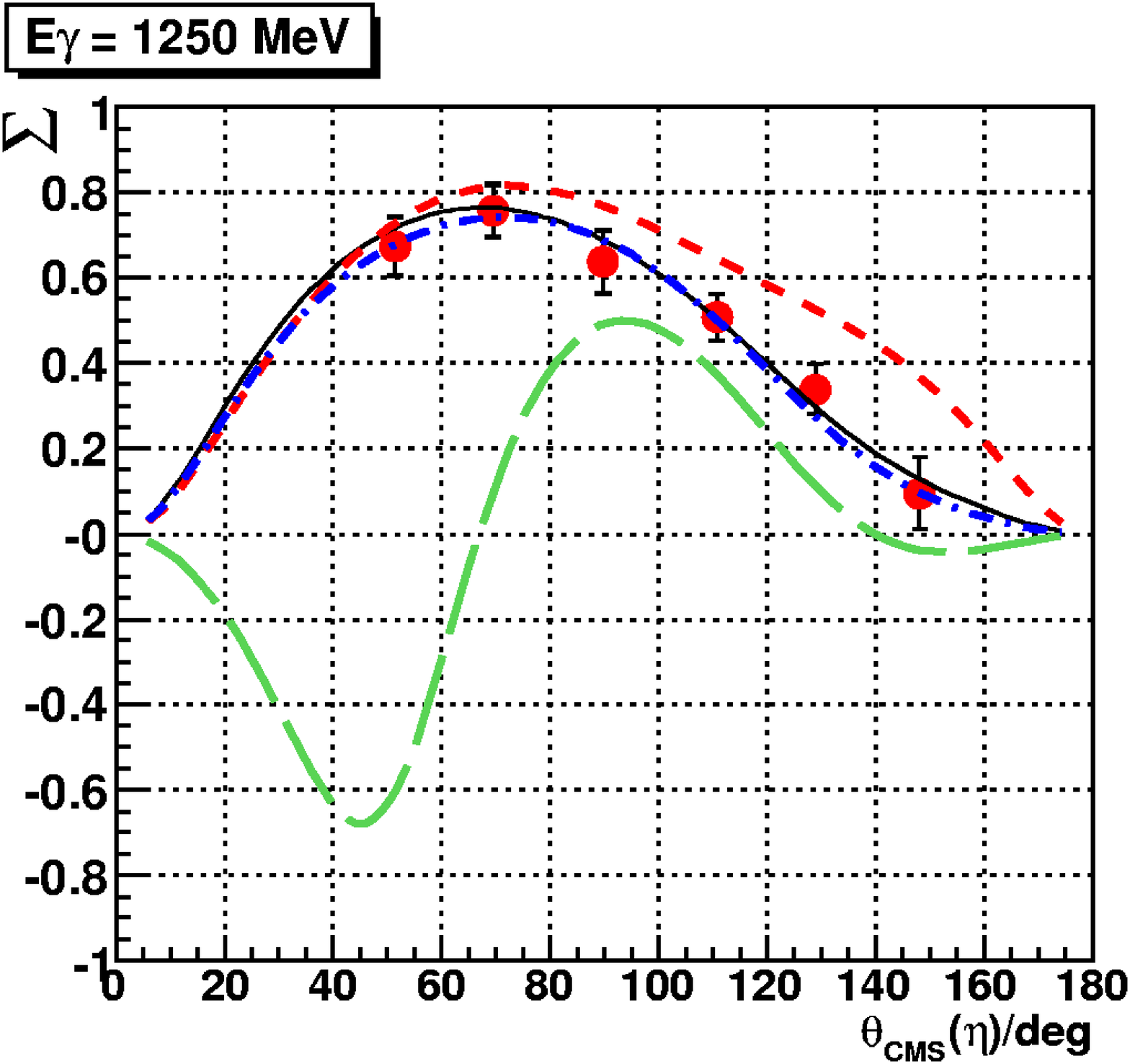,width=7.5cm,angle=0}
\caption{Sensitivity of the \texttt{eta-MAID} and the \texttt{BnGa} 
         calculations to different resonance contributions
         in the energy bin $E_\gamma = (1250 \pm 50)$ MeV.
         Data points are the same as in Fig.\,\ref{fig:Eta_all}.
         Left the {\texttt{eta-MAID}}
         result \cite{CYTD02} is shown, right the 
         \texttt{BnGa} analysis \cite{Anisovich05}.
         The full lines represent the respective full calculations. 
         The broken curves illustrate the impact
         of  ``turning off'' individual resonances:
         Long dashed without $P_{13}(1720)$,
         long dashed-dotted without $P_{11}(1710)$ 
         (no difference to full calculation in \texttt{BnGa} analysis),
         short dashed without $D_{13}(1520)$, and
         short dashed-dotted without $D_{15}(1675)$.}
\label{fig:Sigma-models}       
\end{figure*}
In contrast to \texttt{eta-MAID}, the Bonn--Gatchina analysis
in addition to $\eta N$ also takes the $\pi N$, $K \Lambda$ and $K \Sigma$
coupled channels into account.
To calculate the photon asymmetry, the preliminary high
energy GRAAL data \cite{Kouznetsov02} have already been used in 
the \texttt{BnGa} fit.
This might be the reason for the slightly better description of 
our data.

The overall agreement between data and both models 
seems very satisfactory at first glance.
Closer examination reveals distinct inconsistencies, however. 
While the full model results agree,
the individual resonance contributions differ substantially 
as is illustrated in Fig.\,\ref{fig:Sigma-models}.

Within the energy range considered, the tail of the $S_{11}(1535)$ state 
provides an important contribution to the cross section in both models.
In \texttt{eta-MAID} the $P_{11}(1710)$ is required as well to describe
the cross section, whereas the  
\texttt{BnGa} PWA prefers a strong $P_{13}(1720)$ partial wave.
This also shows up in the photon asymmetry.
The $P_{11}(1710)$ (long dashed-dotted) affects $\Sigma$ in \texttt{eta-MAID},
albeit weakly.
No impact at all is found in the \texttt{BnGa} PWA.
In contrast,
the influence of the $P_{13}(1720)$ (long dashed) on the photon asymmetry
is pronounced only in the \texttt{BnGa} model.
Within \texttt{eta-MAID}, turning off the $P_{13}(1720)$
leaves the photon asymmetry almost unaffected.
Both the $D_{13}(1520)$ (short dashed) and 
$D_{15}(1675)$ (short dashed-dotted) states
have a strong influence on $\Sigma$ within \texttt{eta-MAID}.
Contrary, the $D_{15}(1675)$ remains negligible in the \texttt{BnGa} 
calculation; the $D_{13}(1520)$ has a weak impact but,
compared to \texttt{eta-MAID}, in opposite direction 
(cf. Fig.\,\ref{fig:Sigma-models}).

This unsatisfactory situation can not be resolved from measurements of the
photon asymmetry alone.
Yet, such data provide the necessary basis to be extended with 
double polarisation observables in order to get closer to, or
even accomplish the complete experiment 
in terms of the introductory discussion.

\section{Summary and conclusions}

In summary, we have presented data on the photon beam asymmetry, $\Sigma$, 
in the reaction $\vec\gamma + p \rightarrow p + \eta$.
The continuous $3.2$ GeV ELSA electron beam was used to produce
a linearly polarised tagged photon beam by means of coherent bremsstrahlung
off a diamond crystal, covering a photon energy range 
$E_\gamma = 800 ... 1400$\,MeV with polarisation degrees up to 49\,\%.
A combined setup of the \texttt{Crystal Barrel}
and \texttt{TAPS} detectors enabled high-resolution detection 
of multiple photons, 
important for the clean detection of the 
$2\gamma$ and $3\pi^0$ decays of the $\eta$ meson.
We obtained photon asymmetries in excess of 50\,\% in some 
angular and energy bins.
The results 
are in agreement with a previous measurement by the GRAAL collaboration
in the overlapping energy intervals. 
The \texttt{eta-MAID} model and the 
Bonn--Gatchina partial wave analysis provide a satisfactory
overall description of our data.
In detail, however, there are marked differences with regard to 
the role of individual resonance contributions.
To resolve this problem, further double-polarisation experiments are
indispensable. 
They will be tackled at several laboraties, 
at ELSA within the Collaborative Research Project SFB/TR-16 
with use of the  Bonn polarised solid state target.

\begin{acknowledgement}
We are happy to acknowledge the continuous efforts of the accelerator
crew and operators to provide stable beam conditions.
K. Livingston from Glasgow university deserves a big share
of credit for his invaluable help in setting up the Stonehenge technique
for the crystal alignment.
This work was financially supported by the federal state of 
{\em North Rhine-Westphalia} and the
{\em Deutsche Forschungsgemeinschaft} within the SFB/TR-16.
The Basel group acknowledges support from the
{\em Schweizerischer Nationalfonds}, 
the KVI group from the {\em Stichting voor Fundamenteel Onderzoek der 
Materie} (FOM) and the {\em Nederlandse Organisatie voor Wetenschappelijk 
Onderzoek} (NWO).
\end{acknowledgement}

\section*{Appendix}

The detailed results of the photon asymmetries, $\Sigma$, from the reaction
$\vec\gamma \,p \rightarrow \eta\,p$ are summarised in 
Table\,\ref{tab:results}.
To each value of the photon asymmetry is assigned the corresponding
1-$\sigma$ statistical error and an 1-$\sigma$ estimate of the
total systematical error.  
\begin{table*}
\begin{center}
\begin{tabular}[t]{|c|c|c|c|c||c|c|c|c|c|}
  \multicolumn{5}{l}{energy bin {\bf 850} MeV}
  &\multicolumn{5}{l}{energy bin {\bf 950} MeV}\\
  \hline
  E$_{\gamma}$/MeV & $\theta^{cm}_{\eta}$ &
  $\Sigma$ & $\sigma(\Sigma)_{stat}$ & $\sigma(\Sigma)_{sys}$ & E$_{\gamma}$/MeV & $\theta^{cm}_{\eta}$
&
  $\Sigma$ & $\sigma(\Sigma)_{stat}$ & $\sigma(\Sigma)_{sys}$\\
  \hline
  843.4 & 72.9   & 0.237 & 0.145 & 0.036 & 942.8 & 70.9   & 0.517 & 0.131 & 0.033 \\
  842.6 & 91.2   & 0.382 & 0.087 & 0.020 & 939.7 & 91.9   & 0.546 & 0.095 & 0.045 \\
  846.4 & 109.9  & 0.278 & 0.071 & 0.015 & 941.7 & 110.3  & 0.465 & 0.065 & 0.013 \\
  847.5 & 129.2  & 0.240 & 0.079 & 0.012 & 943.6 & 129.4  & 0.380 & 0.072 & 0.049 \\
  850.5 & 147.9  & 0.114 & 0.116 & 0.031 & 943.8 & 148.2  & 0.184 & 0.095 & 0.023 \\
  \hline
  \multicolumn{10}{l}{}\\
  \multicolumn{5}{l}{energy bin {\bf 1050} MeV}
  &\multicolumn{5}{l}{energy bin {\bf 1150} MeV}\\
  \hline
  E$_{\gamma}$/MeV & $\theta^{cm}_{\eta}$ &
  $\Sigma$ & $\sigma(\Sigma)_{stat}$ & $\sigma(\Sigma)_{sys}$ & E$_{\gamma}$/MeV & $\theta^{cm}_{\eta}$
&
  $\Sigma$ & $\sigma(\Sigma)_{stat}$ & $\sigma(\Sigma)_{sys}$\\
  \hline
  1054.0 & 53.3   & 0.734 & 0.172 & 0.077 & 1154.4 & 52.9   & 0.692 & 0.090 & 0.013 \\
  1051.2 & 70.1   & 0.583 & 0.111 & 0.035 & 1151.5 & 69.8   & 0.731 & 0.072 & 0.019 \\
  1046.0 & 91.4   & 0.559 & 0.108 & 0.032 & 1150.2 & 90.1   & 0.593 & 0.084 & 0.036 \\
  1045.9 & 110.6  & 0.283 & 0.074 & 0.023 & 1151.4 & 110.7  & 0.366 & 0.058 & 0.037 \\
  1043.0 & 129.4  & 0.316 & 0.080 & 0.010 & 1150.0 & 129.0  & 0.165 & 0.064 & 0.014 \\
  1043.3 & 148.4  & 0.261 & 0.107 & 0.019 & 1148.5 & 148.2  & 0.041 & 0.095 & 0.020 \\
  \hline
  \multicolumn{10}{l}{}\\
  \multicolumn{5}{l}{energy bin {\bf 1250} MeV}
  &\multicolumn{5}{l}{energy bin {\bf 1350} MeV}\\
  \hline
  E$_{\gamma}$/MeV & $\theta^{cm}_{\eta}$ &
  $\Sigma$ & $\sigma(\Sigma)_{stat}$ & $\sigma(\Sigma)_{sys}$ & E$_{\gamma}$/MeV & $\theta^{cm}_{\eta}$
&
  $\Sigma$ & $\sigma(\Sigma)_{stat}$ & $\sigma(\Sigma)_{sys}$\\
  \hline
  1251.4 & 51.5   & 0.674 & 0.068 & 0.016 & 1344.7 & 50.7   & 0.687 & 0.083 & 0.027 \\
  1249.3 & 69.4   & 0.758 & 0.060 & 0.017 & 1343.5 & 69.4   & 0.774 & 0.075 & 0.069 \\
  1249.0 & 89.8   & 0.638 & 0.073 & 0.012 & 1342.4 & 89.4   & 0.680 & 0.102 & 0.043 \\
  1249.5 & 110.7  & 0.508 & 0.053 & 0.021 & 1342.6 & 111.3  & 0.479 & 0.070 & 0.026 \\
  1249.5 & 129.0  & 0.338 & 0.058 & 0.010 & 1343.4 & 129.0  & 0.290 & 0.075 & 0.021 \\
  1250.0 & 148.2  & 0.095 & 0.085 & 0.076 & 1343.1 & 147.9  & 0.149 & 0.119 & 0.089 \\
  \hline
\end{tabular}
\caption{Photon asymmetries for the reaction $\vec{\gamma}p\rightarrow \eta p$.
         Angles are given in degrees. Energy-bin widths are $\pm 50$ MeV.}
\label{tab:results}
\end{center}
\end{table*}


\end{document}